\documentclass{aa}  

\usepackage{natbib}
\usepackage{graphicx}
\usepackage{txfonts}

\newcommand{\gammaSI}{{\,J\,m$^{-2}$\,s$^{-0.5}$\,K$^{-1}$}}
\newcommand{\dd}{\mathrm{d}}

\begin{document} 

  \title{Asteroid shapes and thermal properties from combined optical and mid-infrared photometry inversion}

   \author{J. \v{D}urech\inst{1} \and 
          M. Delbo'\inst{2} \and
	  B. Carry\inst{2} \and
	  J. Hanu\v{s}\inst{1} \and
	  V. Al\'i-Lagoa\inst{2,3}
          }

   \institute{Astronomical Institute, Faculty of Mathematics and Physics, Charles University, V Hole\v{s}ovi\v{c}k\'ach 2, 180\,00 Prague 8, Czech Republic\\
     \email{durech@sirrah.troja.mff.cuni.cz}
     \and
     Universit\'e C{\^o}te d'Azur, Observatoire de la C{\^o}te d'Azur, CNRS, Laboratoire Lagrange, France
     \and
     Max-Planck-Institut f\"{u}r extraterrestrische Physik, Giessenbachstra{\ss}e, Postfach 1312, 85741 Garching, Germany
   }

   \date{Received ???; accepted ???}

  \abstract
  % context heading (optional)
   {Optical lightcurves can be used for the shape and spin
     reconstruction of asteroids. Due to unknown albedo, these models
     are scale-free. When thermal infrared data are available, they
     can be used for scaling the shape models and for deriving
     thermophysical properties of the surface by applying a
     thermophysical model.} 
  % aims heading (mandatory)
   {We introduce a new method of simultaneous inversion of optical and thermal infrared data that allows the size 
of an asteroid to be derived along with its shape and spin state.}
  % methods heading (mandatory)
   {The method optimizes all relevant parameters (shape and its size, spin state, light-scattering properties, 
thermal inertia, surface roughness) by gradient-based optimization. The thermal emission is computed by solving the 
1-D heat diffusion equation. Calibrated optical photometry and thermal fluxes at different wavelengths are needed as input data.} 
  % results heading (mandatory)
   {We demonstrate the reliability and test the accuracy of the method on selected targets with different amount and 
quality of data. Our results in general agree with those obtained by independent methods.}
  % conclusions heading (optional)
   {Combining optical and thermal data into one inversion method opens a new possibility for processing photometry 
from large optical sky surveys with the data from WISE. It also provides more realistic estimates of errors of thermophysical parameters.}


   \maketitle

\section{Introduction}

Optical and thermal infrared disk-integrated radiation of asteroids is routinely used for the determination of their physical properties 
\citep[][and references therein]{Kaa.ea:02c, Har.Lag:02, Dur.ea:15b, Del.ea:15}. The reflected sunlight at optical wavelengths serves as 
the first guess for the size of the asteroids (from the brightness) and taxonomic type (color information). The periodic variations of 
the brightness due to rotation carry information about the shape and spin state of the asteroid. A mathematically robust and reliable 
method of reconstruction of asteroid's shape and spin state from disk-integrated reflected light was developed by \cite{Kaa.Tor:01} and
\cite{Kaa.ea:01}. The method has been successfully applied to hundreds of asteroids \citep[][for example]{Dur.ea:09, Han.ea:11, Han.ea:16, Mar.ea:11, Dur.ea:16} 
and its results were confirmed by independent methods \citep{Kaa.ea:05, Mar.ea:06, Kel.ea:10, Dur.ea:11}. Due to the ill-posedness of the 
general inverse problem in case of disk-integrated data, the shape is usually modelled as convex, because this approach guarantees uniqueness 
of the model. The output of the lightcurve inversion method is a shape model represented by a convex polyhedron that approximates the real 
nonconvex shape of an asteroid. Because of the ambiguity between size $D$ and geometric albedo $p_\text{V}$, the models are scale free. 
The visible flux is proportional to $D^2 p_\text{V}$ and in principle $0.01 \lesssim p_\text{V} \lesssim 1.0$ \citep{Mas.ea:11}, so size 
estimation from the visible brightness has a very large uncertainty. The models can be set to scale by disk-resolved data \citep{Han.ea:13a}, 
stellar occultation silhouettes \citep{Dur.ea:11}, or thermal infrared data \citep{Han.ea:15}.
 
The transition from purely reflected light to purely thermal emission is continuous and the transition zone where the detected 
flux is a mixture of reflected solar radiation and thermal emission depends on the heliocentric distance, albedo, and thermal 
properties of the surface. For asteroids in the main-belt it is around 3--5\,$\mu$m \citep{Del.Har:02, Har.Lag:02}. For longer 
wavelengths, the flux can be treated as pure thermal emission. Measurements of this flux can be used for direct estimation of 
the size and sophisticated thermal models have been developed to reveal thermal properties of the surface from measurements of 
thermal emission at different wavelengths.

The simplest approach to thermal modelling that is often used when no
information about the shape is available is to assume that the shape
is a sphere. Then the standard thermal model \citep[STM,][]{Leb.ea:86}
or the near-Earth asteroid thermal model \citep[NEATM,][]{Har:98} are
used. When interpreting thermal data by a thermophysical model (TPM),
the shape and spin state of the asteroid has to be known to be able to
compute the viewing and illumination geometry for each facet on the
surface (and shadowing in case of a nonconvex model). Usually, the 1-D
heat diffusion problem is solved in the subsurface layers. There are
many different TPM codes, they differ mainly in the way they deal with
surface roughness \citep[for a review, see][]{Del.ea:15}. Shape models
are usually reconstructed from other data sources like photometry
\citep{Kaa.ea:02c}, radar echos \citep{Ben.ea:15}, or high-angular-resolution
imaging \citep{Car.ea:10}.
Although this approach in general works and provides thermophysical parameters, the main caveat here is that the shape and 
spin state are taken as a~priori and their uncertainties are not taken into account. This may lead to underestimation of 
errors of the derived parameters or even erroneous results \citep[see][for example]{Roz.Gre:14, Han.ea:15}.

To overcome the limitation of a two-step approach where first the
shape/spin model is created and then thermal data are fitted, we have
developed a new algorithm that allows for simultaneous optimization of all
relevant parameters. We call it convex inversion TPM (CITPM) and we
describe the algorithm in Sect.~\ref{sec:method} and show how this
method works for some test asteroids in Sect.~\ref{sec:results}. 

\section{Combined inversion of optical and thermal infrared data}
\label{sec:method}

Our new code joins two widely used and well tested methods: (i) the lightcurve inversion of \cite{Kaa.ea:01} and (ii) the 
thermophysical model of \cite{Lag:96a, Lag:97, Lag:98}. We use the convex approach, which enables us to work in the Gaussian 
image representation: The convex shape is represented by areas of surface facets and corresponding normals. The normals are 
fixed while the areas are optimized to get the best agreement between the visible light and the thermal infrared fluxes 
calculated by the model and the observed fluxes. Moreover, the distribution of individual areas is parametrized by spherical 
harmonics (usually of order and degree of six to eight). The polyhedral representation of the shape is then reconstructed by 
the Minkowski iteration \citep{Kaa.Tor:01}. The spin vector is parametrized by the direction of the spin axis in ecliptic 
coordinates $(\lambda, \beta)$ and the sidereal rotation period $P$. Together with the initial orientation $\varphi_0$ at 
epoch JD$_0$, these parameters uniquely define the orientation of the asteroid in the ecliptic coordinate frame \citep{Dur.ea:10}. 
With known positions of the Sun and Earth with respect to the asteroid, the illumination and viewing geometry can be computed for 
each facet. As we work with convex shapes only, there is no global shadowing by large-scale topography in our model, although the 
generalization of the problem with nonconvex shapes is straightforward, similarly as in the case of lightcurve inversion. 

For computing the brightness of an asteroid in visible light, we use Hapke's model with shadowing \citep{Hap:81, Hap:84, Hap:86}. 
This model has five parameters: the average particle single-scattering albedo $\varpi_0$, the asymmetry factor $g$, the width $h$ 
and amplitude $B_0$ of the opposition effect, and the mean surface slope $\bar{\theta}$ . We compute Hapke's bidirectional reflectance 
$r(i,e,\alpha)$ for each surface element, where $i$ is the incidence angle, $e$ is the emission angle, and $\alpha$ is the phase angle. 
The total flux scattered towards an observer is computed as a sum of contributions from all visible and illuminated facets. 

In standard TPM methods, the size of the asteroid $D$ in kilometers and its geometric albedo $p_\text{V}$ are connected via Bond albedo 
$A_\text{B}$, phase integral $q$, and absolute magnitude $H_0$ with formulas \citep[see][for example]{Har.Lag:02}:
\begin{equation}
 D = \frac{1329}{\sqrt{p_\text{V}}}\, 10^{-\frac{H_0}{5}}, \qquad A_\text{B} = q p_\text{V}\,.
\end{equation}
However, Bond albedo $A_\text{B}$ as well as geometric visible albedo $p_\text{V}$ are not material properties and they are unambiguously 
defined only for a sphere. So instead of this traditional approach, we use a self-consistent model, where the set of Hapke's parameters 
is also used to compute the total amount of light scattered to the upper hemisphere, which defines the hemispherical albedo $A_\text{h}$ 
\citep[the ratio of power scattered into the upper hemisphere by a unit area of the surface to the collimated power incident on the unit 
surface area,][]{Hap:12} needed for computing the energy balance between incoming, emitted, and reflected flux. This hemispherical albedo 
is dependent on the angle of incidence (it is different for each surface element) and it is at each step computed by numerically evaluating the integral
\begin{equation}
 \label{eq:hemispherical_albedo}
 A_\text{h}(i) = \frac{1}{\mu_0} \int_\Omega r(i,e,\alpha)\, \mu \, \dd\Omega\,,
\end{equation}
where $\mu = \cos e$, $\mu_0 = \cos i$, and the integration region $\Omega$ is over the upper hemisphere. Because the above defined 
quantities are in general dependent on the wavelength $\lambda$, we need bolometric hemispherical albedo $A_\text{bol}$ that is 
the average of the wavelength-dependent hemispherical albedo $A_\text{h}(\lambda)$ weighted by the spectral irradiance of the Sun $J_\text{S}(\lambda)$:
\begin{equation}
 \label{eq:bolometric_albedo}
 A_\text{bol} = \frac{\int_0^\infty A_\text{h}(\lambda) J_\text{s}(\lambda)\, \dd\lambda}{\int_0^\infty J_\text{s}(\lambda)\, \dd\lambda} \,.
\end{equation}
In practice, we approximate the integrals by sums. The dependence of $A_\text{h}$ on $\lambda$ is not know neither from theory nor 
measurements, so we assume that it is the same as the dependence of the physical (geometric) albedo $p(\lambda)$ that can be 
estimated from reflectance curves. If asteroid's taxonomy is known, we take the spectrum of that class in the Bus-DeMeo taxonomy 
\citep{DeM.ea:09}, and extrapolate the unknown part outside the 0.45--2.45\,$\mu$m interval with flat reflectance. We then multiply 
this spectrum by the spectrum of the Sun, taken from \cite{Gue:04}. If the taxonomic class is unknown, we assume flat 
reflectance, i.e., $A_\text{h}(\lambda) = \text{constant}$.
 
Both Bond and geometric albedos can be computed from Hapke's parameters but they are not used in our model directly. 
We also do not need the formalism of HG system \citep{Bow.ea:89}. Similarly as geometric albedo, the $H_0$ value is 
properly defined only for a sphere, for a real irregular asteroid it depends on the aspect angle. So instead of using 
$H_0$ and $p_\text{V}$, we use directly the calibrated magnitudes on the phase curve. With disk-integrated photometry 
and a limited coverage of phase angles, it is usually not possible to uniquely determine the Hapke's parameters. In such 
cases, they (or a subset of them) can be fixed at some typical values \citep[see Table 6 of][]{Li.ea:15}.

Parameters of the thermophysical model are the thermal inertia $\Gamma$, the fraction of surface covered by craters 
$\rho_\text{c}$ and their opening angle $\gamma_\text{c}$ \citep{Lag:98}. The roughness parameter $\bar\theta$ of 
Hapke's model can be set to the value corresponding to craters or set to a different value, which would mean that there 
are two values of roughness in the model, one for optical and one for infrared wavelengths. Usually, we use just one value,
 which makes the model simpler and also more self-consistent.

To find the best-fitting parameters, we use the Levenberg-Marquardt algorithm. The parameters are optimized to give the 
lowest value for the total $\chi^2$ that is composed from the visual and infrared part weighted by $w$:
\begin{equation}
 \label{eq:chisq}
 \chi^2_\text{total} = \chi^2_\text{VIS} + w \chi^2_\text{IR}\,.
\end{equation}
The visual part is computed as a sum of squares of differences between the observed flux $F_\text{VIS, obs}$ and the modelled 
flux $F_\text{VIS, model}$ over individual data points $i$ weighted by the measurement errors $\sigma_i$:
\begin{equation}
 \chi^2_\text{VIS} = \sum_i \left( \frac{F_\text{VIS, obs}^{(i)} - F_\text{VIS, model}^{(i)}}{\sigma_i} \right)^2\,.
\end{equation}
The model flux is computed as a sum over all illuminated and visible surface elements $k$:
\begin{equation}
 F_\text{VIS, model} = \frac{F_\text{in}}{\Delta^2} \sum_k   r_k\,\mu_k\,\delta\sigma_k\,,
\end{equation}
where $F_\text{in}$ is the incident solar irradiance, $\Delta$ is the distance between the observer and the asteroid, $r$ 
is Hapke's bidirectional reflectance, $\mu$ is cosine of the emission angle, and $\delta\sigma$ is area of the surface element. 
If the photometry is not calibrated and we have only relative lightcurves, the corresponding $\chi^2$ part is computed such 
that only the relative fluxes are compared:
\begin{equation}
 \chi^2_\text{rel} = \sum_j \sum_i \left( \frac{F_\text{VIS, obs}^{(j)(i)}}{\bar{F}_\text{VIS, obs}^{(j)}} -  \frac{F_\text{VIS, model}^{(j)(i)}}{\bar{F}_\text{VIS, model}^{(j)}}\right)^2\,,
\end{equation}
where $j$ is an index for lightcurves, $i$ is an index for individual points of a lightcurve, and $\bar{F}^{(j)}$ is the mean brightness of the $j-$th lightcurve.
 
Similarly, for the IR part of the $\chi^2$ we compute
\begin{equation}
 \chi^2_\text{IR} = \sum_i \left( \frac{F_\text{IR, obs}^{(i)} - F_\text{IR, model}^{(i)}}{\sigma_i} \right)^2\,,
\end{equation}
where we compare the observed thermal flux $F_\text{IR, obs}$ with the modelled flux $F_\text{IR, model}$ at some 
wavelength $\lambda$. The disk-integrated flux is a sum of contributions from all surface elements that are visible to the observer:
\begin{equation}
 F_\text{IR, model} = \frac{\epsilon}{\Delta^2} \sum_k B(\lambda, T_k, \rho_\text{c}, \gamma_\text{c})\,\mu_k\,\delta\sigma_k\,,
\end{equation}
where $\epsilon$ is the emissivity (assumed to be independent on $\lambda$), and $B$ is radiance of a surface element 
at temperature $T$ that depends on the wavelength $\lambda$ (blackbody radiation) and includes also a model for 
macroscopic roughness parametrized by $\rho_\text{c}$ and $\gamma_\text{c}$. To compute $B$, we have to solve the heat 
diffusion equation for each surface element and compute $T_k$. Assuming that the material properties do not depend on temperature, we solve
\begin{equation}
 \label{eq:HTE}
 \rho C \frac{\partial T}{\partial t} = \kappa \frac{\partial^2 T}{\partial z^2}\,,
\end{equation}
where the density $\rho$, heat capacity $C$, and thermal conductivity $\kappa$ are combined into a single parameter thermal 
inertia $\Gamma = \sqrt{\rho C \kappa}$ \citep{Spe.ea:89, Lag:96a}. Instead of the real subsurface depth $z$, the problem 
is then solved in the units of the thermal skin depth, which is related to the rotation period and the thermal properties 
of the material. The boundary condition at the surface ($z = 0$) is the equation for energy balance:
\begin{equation}
 (1 - A_\text{bol}) \frac{F_\sun}{r^2} \mu_0 = \epsilon \sigma T^4 - \kappa \frac{\partial T}{\partial z}\,,
\end{equation}
where the left side of the equation is the energy absorbed by the surface ($F_\sun$ is solar irradiance at 1\,au, $r$ is the 
distance from the Sun, and $\mu_0$ is cosine of the incidence angle) and the right side is the radiated flux ($\sigma$ is the 
Stefan-Boltzmann constant) minus the heat transported inside the body. The inner boundary condition is
\begin{equation}
 \left. \frac{\partial T}{\partial z}\right |_{z \rightarrow \infty} = 0\,,
\end{equation}
where $z \rightarrow \infty$ in practice means few skin depths. Eq.~(\ref{eq:HTE}) is solved by the simplest explicit 
method with typically tens of subsurface layers.

The weight $w$ in eq.~(\ref{eq:chisq}) is set such that there is a balance between the level
of fit to lightcurves and thermal data. Objectively, the optimum value
can be found with the method proposed by \cite{Kaa:11} with the
so-called maximum compatibility estimate, which corresponds to the
maximum likelihood or maximum a~posteriori estimates in the case of a
single data mode. 

Because of strong correlation between the thermal inertia $\Gamma$ and the surface roughness, the parameters describing 
the surface fraction covered by craters $\rho_\text{c}$ and their opening angle $\gamma_\text{c}$ and likewise the 
parameter $\bar\theta$ in Hapke's model are the only three parameters that are not optimized. They are held fixed and their 
best values are found by running the optimization many times with different combination of these parameters. The dependence 
of $\chi^2_\text{total}$ on $\Gamma$ shows usually only one minimum (sharp or shallow depending on the amount and quality 
of thermal data), which makes convergence in $\Gamma$ robust and the gradient-based optimization converges to the best 
value of $\Gamma$ even if started far from it. The emissivity $\epsilon$ is held fixed. Its value can be set to anything 
between 0 and 1, but in all tests in the following section we used the ``standard'' value $\epsilon = 0.9$. This is a 
typical value of the emissivity of meteorites and of the minerals included in meteorites obtained in the lab at wavelengths 
around 10\,$\mu$m. This value can significantly change at longer wavelengths, see \cite{Del.ea:15} for discussion.

\section{Testing the method on selected targets}
\label{sec:results}

We tested our method on selected targets, each representing a typical amount of data. First, we used asteroid 
(21)~Lutetia for which there are plenty of photometry and thermal infrared data available and the shape is known 
from Rosetta spacecraft fly-by. So it is an ideal target for comparing the results from our inversion with ground 
truth from Rosetta. Another test asteroid with known shape is (2867)~\v{S}teins, but in this case thermal data are 
scarce, so this example should show us the limits of the method. Then we have selected asteroid (306)~Unitas -- 
a typical example of an asteroid with enough lightcurves and some thermal data from IRAS and WISE satellites. As the 
last example, we show on asteroid (220)~Stephania how thermal data in combination with sparse optical photometry can 
lead to a unique model -- this is perhaps the most important aspect of the new method that can lead to production 
of new asteroid models in the future.

\subsection{Data}

The success of the method is based on the assumption that we have
accurately calibrated absolute photometry of an asteroid covering a
wide interval of phase angles -- to be able to use Hapke's photometric
model and derive the hemispherical albedo. Because most of the
available lightcurves are only relative, we used also the Lowell
Observatory photometric database \citep{Osz.ea:11, Bow.ea:14} as the
source of absolutely calibrated photometry in V filter. The only
asteroid for which we had enough reliable absolutely calibrated dense
lightcurves was (21)~Lutetia. As regards thermal data, we used IRAS
catalogue \citep{Ted.ea:04}, WISE data \citep{Wri.ea:10, Mai.ea:11b},
and also Spitzer and Herschel observations for (21)~Lutetia. The
number of dense lightcurves (usually relative), sparse data points
(calibrated), and IR data points for all targets is listed in
Table~\ref{tab:data}. The comparison between CITPM models and
independent models is shown in Table~\ref{tab:models} in terms of
derived physical parameters. 

\subsection{(21) Lutetia}

Because for this asteroid there are lot of lightcurves that well
define the convex shape, we do not expect any significant improvement
of the model with respect to the two-step method. However, Lutetia
with its abundant photometric and infrared data set is a good test
case for the new algorithm. 

A convex model of Lutetia was derived by \cite{Tor.ea:03} and later
\cite{Car.ea:10b} created a nonconvex model that was confirmed by the
Rosetta fly-by \citep{Car.ea:12}. The detailed shape model of Lutetia reconstructed from
the fly-by imaging \citep{Sie.ea:11} can serve as a ground-truth
comparison with our model (although part of the surface was not seen
by Rosetta and was reconstructed from lightcurves). There are also
abundant optical and thermal infrared data for this asteroid -- 59
lightcurves observed between 1962 and 2010, out of which 20 are
calibrated in V filter, and thermal infrared data \citep{ORo.ea:12}
from 7.87 (Spitzer) to 160\,$\mu$m (Herschel PACS). We have not
included the Herschel SPIRE data observed at 250, 350, and
500\,$\mu$m, because at these long wavelengths the emissivity
assumption of $\epsilon = 0.9$ is no longer valid \citep[see the
  discussion in][for example]{Del.ea:15}.  
%ok with modification for epsilon

\begin{figure}
   \begin{center}
    \includegraphics[trim=0 7cm 0 0, clip, width=\columnwidth]{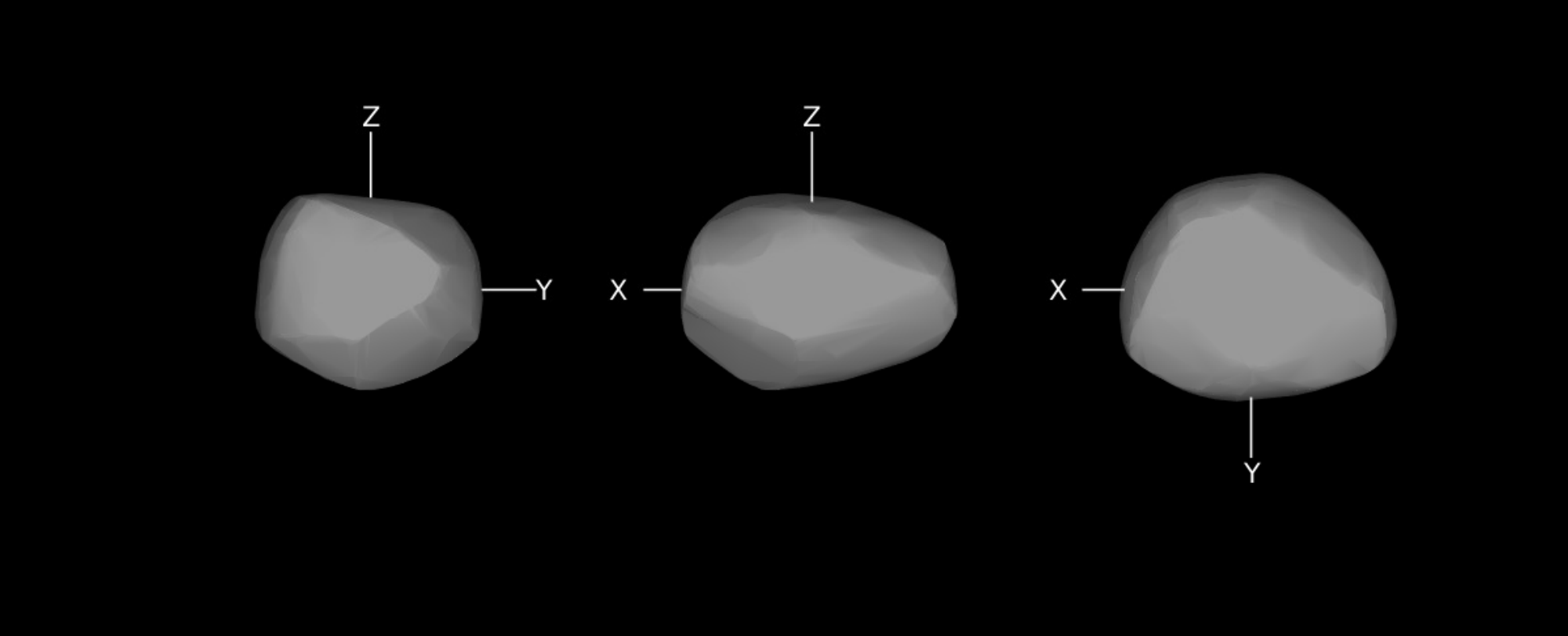}
    \includegraphics[trim=0 7cm 0 0, clip, width=\columnwidth]{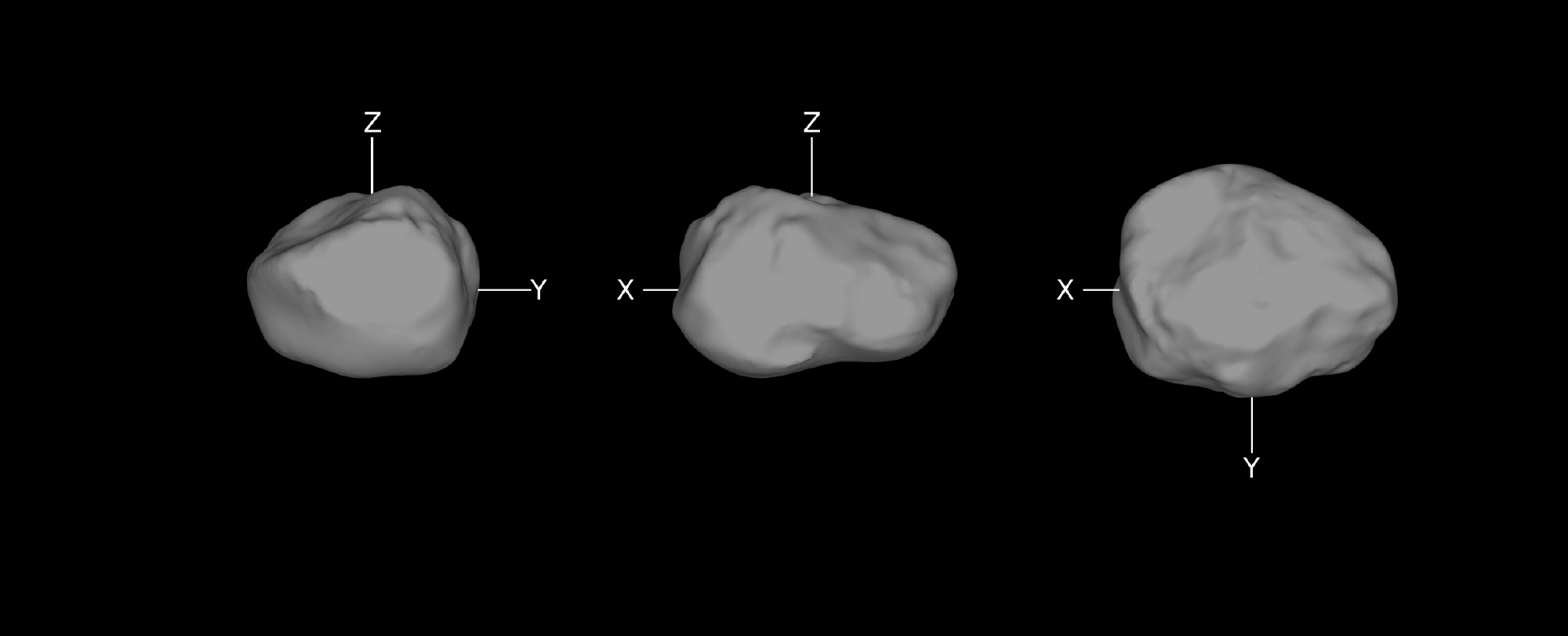}
    \caption{Comparison between the shape of (21) Lutetia reconstructed by our method (top) and that reconstructed by \cite{Sie.ea:11} from Rosetta fly-by images (bottom).}
    \label{fig:Lutetia}
   \end{center}
  \end{figure}

The comparison between our model reconstructed from disk-integrated lightcurves and thermal data and that 
reconstructed by \cite{Sie.ea:11} from Rosetta fly-by images is shown in Fig.~\ref{fig:Lutetia}. The 
volume-equivalent diameter of Rosetta-based model is $98 \pm 2$\,km, while our model has diameter of $101 \pm 4$\,km. 
Our thermal inertia of 30--50\,\gammaSI\ is significantly higher than the value $\sim 5$\,\gammaSI\ of \cite{ORo.ea:12}, 
at the boundary of interval $< 30$\,\gammaSI\ given by \cite{Kei.ea:12}, and consistent with the value of $\leq 100$\,\gammaSI\ 
of \cite{Mul.ea:06} derived from IRAS and IRTF (InfraRed Telescope Facility) data. The discrepancy between our value and 
those by \cite{ORo.ea:12} and \cite{Kei.ea:12} might be partly caused by the fact that we used only wavelengths $\leq 160\,\mu$m, 
while the lower values of thermal inertia were derived from data sets containing also sub-millimeter and millimeter wavelengths. 
Longer wavelengths ``see'' deeper in the subsurface, which is colder than the surface layers. Since $\Gamma$ depends on the 
temperature, we expect to see $\Gamma$ decreasing with depth, provided the density and packing of the regolith is independent on 
the depth, which might not be the case. For example, on the Moon the regolith density increases with depth \citep{Vas.ea:12}. 
\cite{Har.Dru:16} claim to see the same effect on asteroids. 

\subsection{(306) Unitas}

\begin{figure}
   \begin{center}
    \includegraphics[trim=0 5cm 0 0, clip, width=\columnwidth]{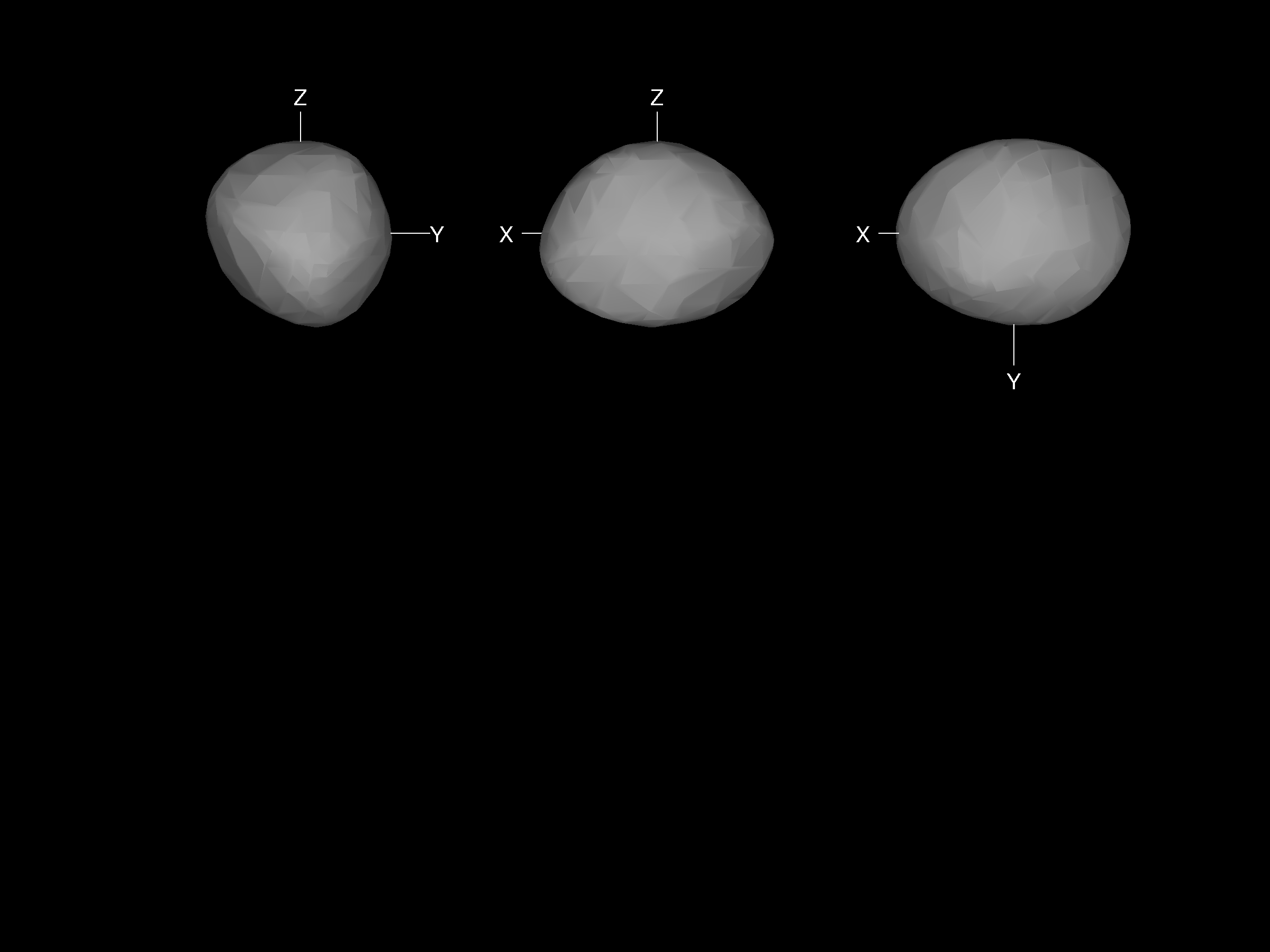}
    \includegraphics[trim=0 5cm 0 0, clip, width=\columnwidth]{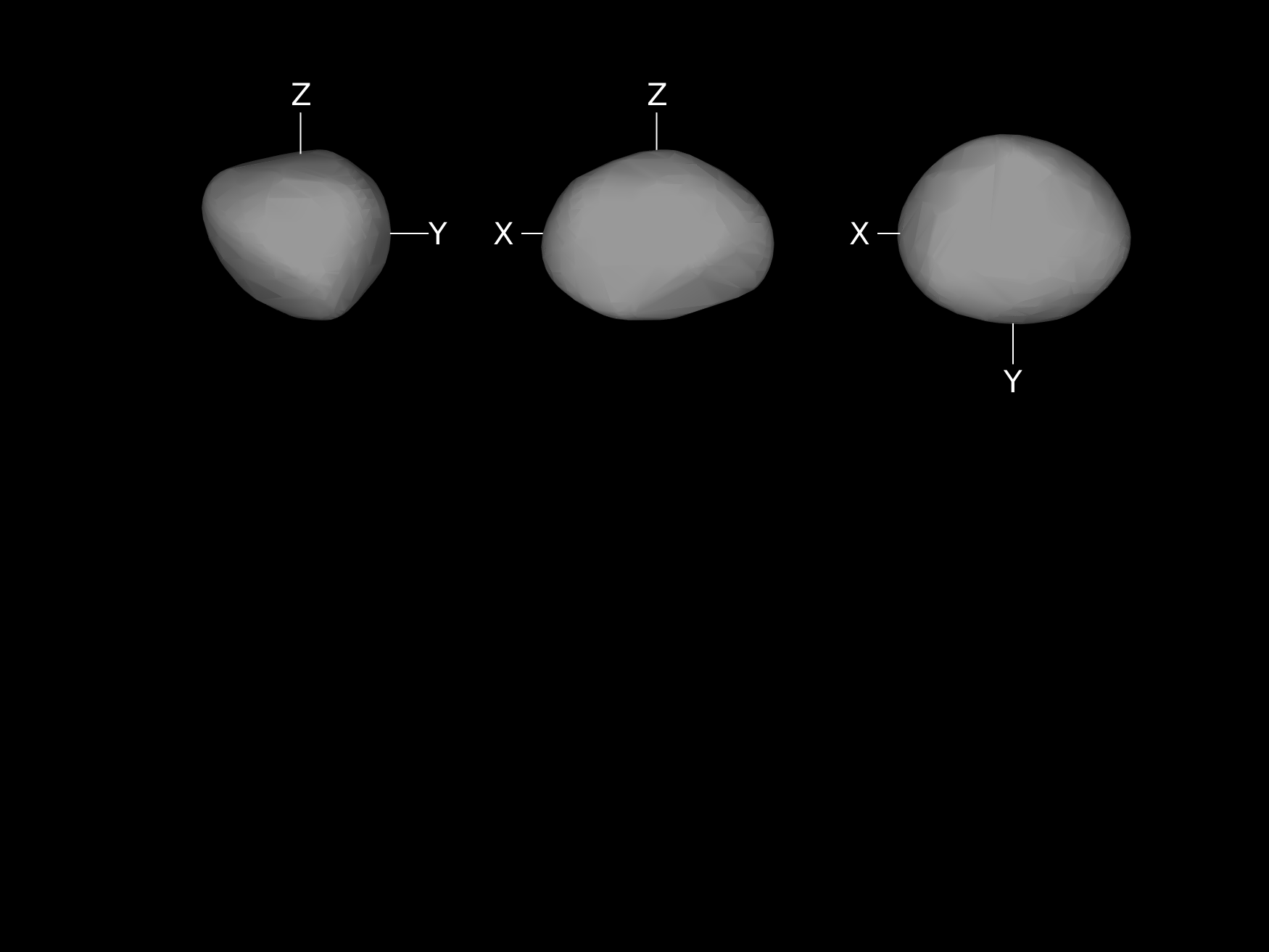}
    \caption{Comparison between the new shape of (306) Unitas
      reconstructed from thermal and optical data (top) and that
      reconstructed only from dense optical lightcurves by
      \cite{Dur.ea:07} (bottom).}
    \label{fig:Unitas_shape}
   \end{center}
  \end{figure}

Unitas is a typical example of a main belt asteroid for which there are some data from both IRAS and WISE 
(see Tab.~\ref{tab:data}). The first convex model was published by \cite{Dur.ea:07} with pole ambiguity that
 was later resolved with IR data by \cite{Del.Tan:09} and confirmed with a fit to stellar occultation data by 
\cite{Dur.ea:11}. The new CITPM shape model obtained from combined inversion of optical lightcurves and thermal 
data has the pole direction $(\lambda, \beta) = (82^\circ, -30^\circ)$ and the shape similar (Fig.~\ref{fig:Unitas_shape}) 
to the old model  with pole direction of $(79^\circ, -35^\circ)$. The size of the formally best-fitting model 
(derived independently on the occultation) is $48.7 \pm 0.7$\,km, which is very close to the size $49 \pm 5$\,km 
obtained by scaling the lightcurve-based model to occultation chords. For comparison, the size from the IRAS 
observation is $46.7 \pm 2.3$\,km \citep{Ted.ea:04}, from WISE $51.6 \pm 6.3$\,km \citep{Mas.ea:11}, and from 
AKARI $46.2 \pm 0.6$\,km \citep{Usu.ea:11}. 

\begin{figure}
   \begin{center}
    \includegraphics[width=\columnwidth]{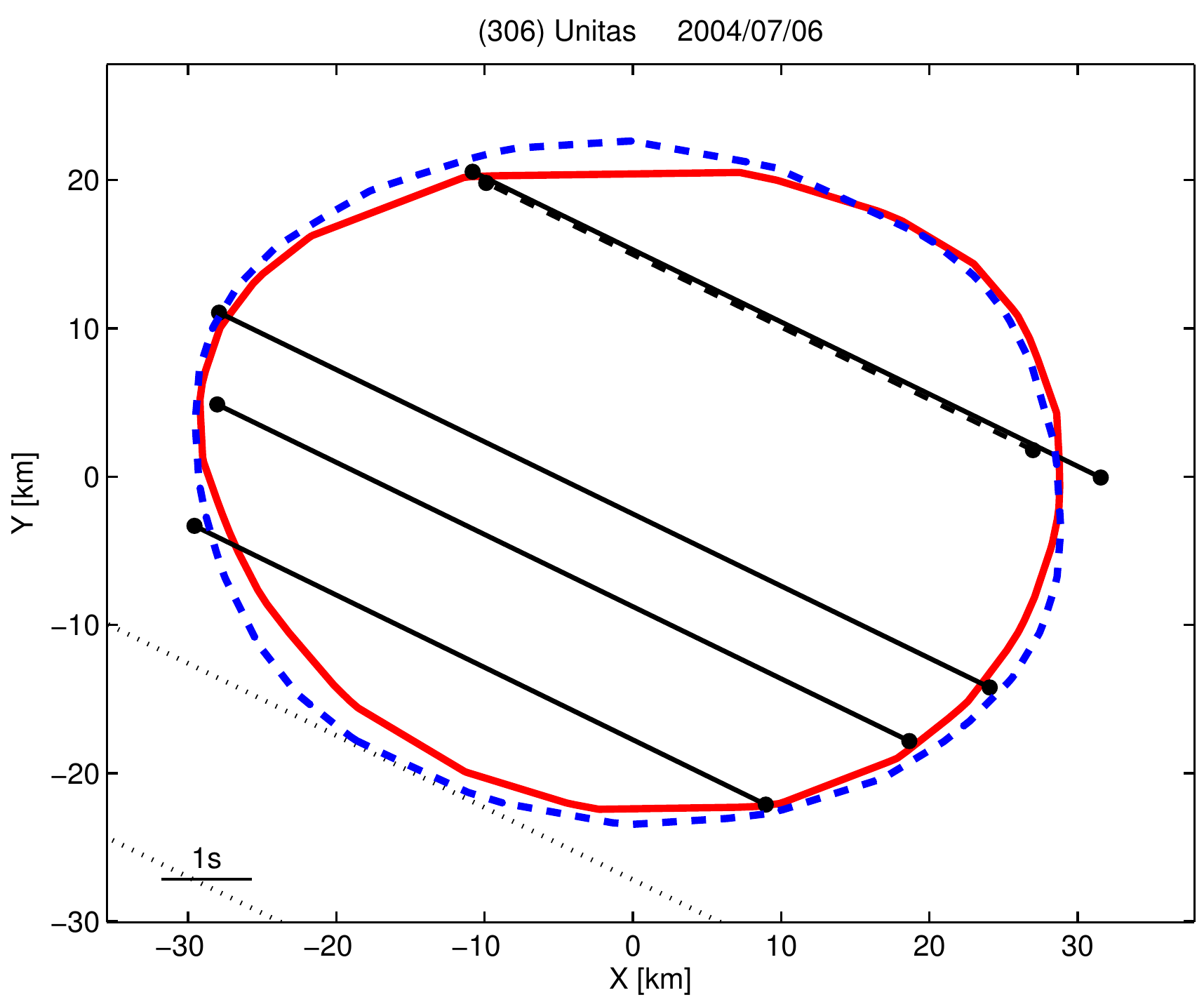}
    \caption{The plane-of-the-sky silhouettes of (306) Unitas for the
      time of the occultation. The red solid silhouette corresponds to
      our model, the blue dashed one corresponds to the model from
      %minor edit
      \cite{Dur.ea:07}. The black lines represent chords reconstructed
      from timings and positions of individual observers. The dashed
      chord represents a visual observation with uncertain timing, the
      dotted lines are a negative observations with no occultation
      detected. Occultation data were taken form \cite{Dun.ea:16}} 
    \label{fig:Unitas_occ}
   \end{center}
  \end{figure}

\begin{figure}
   \begin{center}
    \includegraphics[width=\columnwidth]{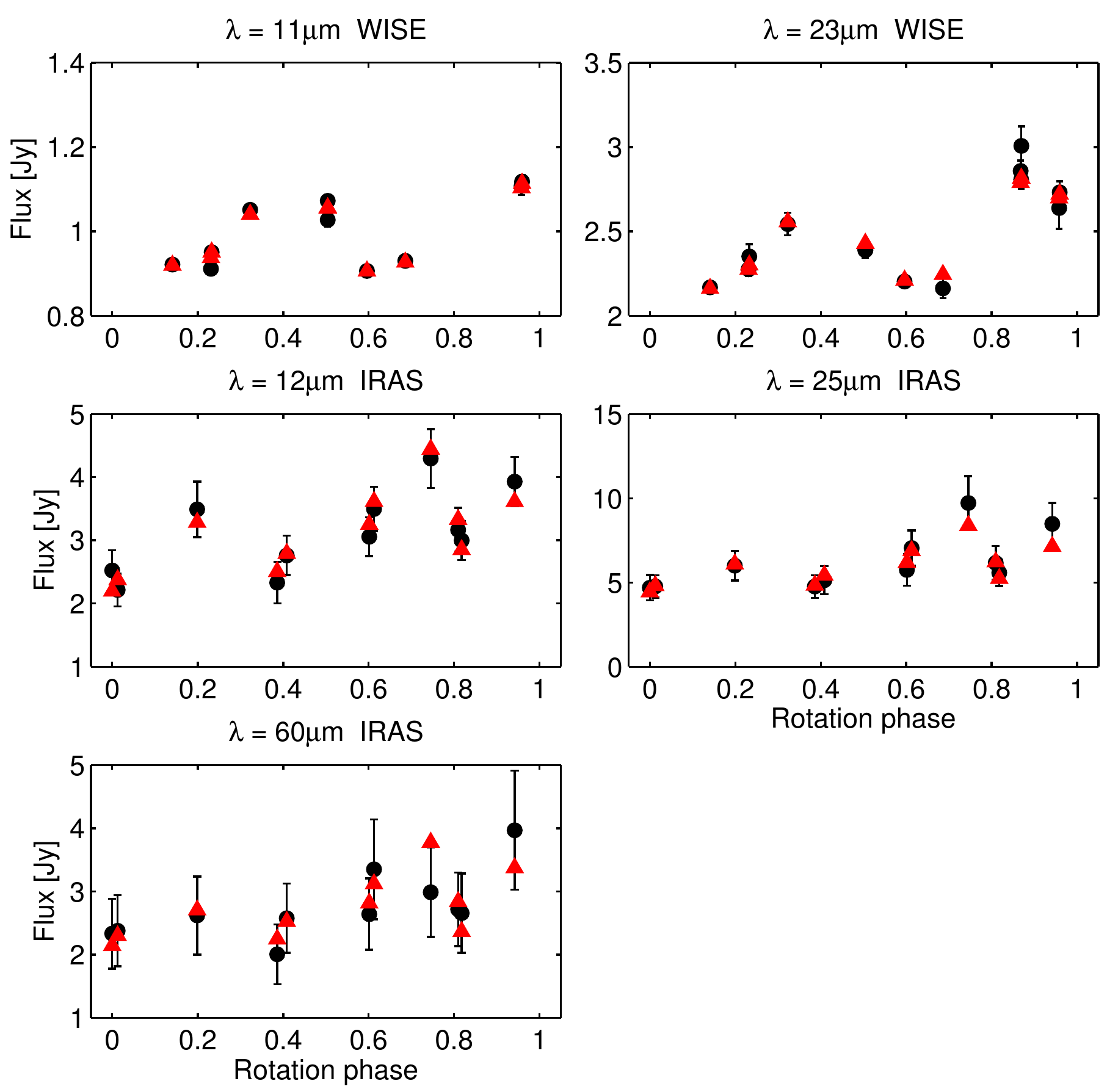}
    \caption{Measured IR fluxes (black circles) of (306) Unitas and the model fluxes for our model (red triangles).}
    \label{fig:Unitas_flux}
   \end{center}
  \end{figure}

The comparison between the model silhouette and the fit to occultation is shown in Fig.~\ref{fig:Unitas_occ}, 
where the red contour corresponds to our new model (no scaling, rms residual 1.7\,km) and the blue one corresponds 
the the model of \cite{Dur.ea:07} (rms residual 1.9\,km) that was scaled to provide the best fit to the chords. The 
fit between the model and observations is formally very good (Fig.~\ref{fig:Unitas_flux}) with reduced $\chi^2_\text{IR}$ 
as low as $\sim 0.7$ for thermal inertia in the range 10--100\gammaSI. \cite{Del.Tan:09} applied TPM on only IRAS data 
and derived systematically higher thermal inertia 100--260\gammaSI\ and also size 55--56\,km -- this demonstrates the 
fact that with a fixed shape and spin as an input for TPM, the uncertainties of thermophysical parameters are often 
underestimated	 \citep[see also][]{Han.ea:15}.

As can be seen in Fig.~\ref{fig:Unitas_shape}, adding IR data changed the shape model only slightly, but they allowed 
correct scaling of the size that can be independently checked by occultation in Fig.~\ref{fig:Unitas_occ}.

\subsection{(2867) \v{S}teins}

\begin{figure}
   \begin{center}
    \includegraphics[trim=0 5cms 0 0, clip, width=\columnwidth]{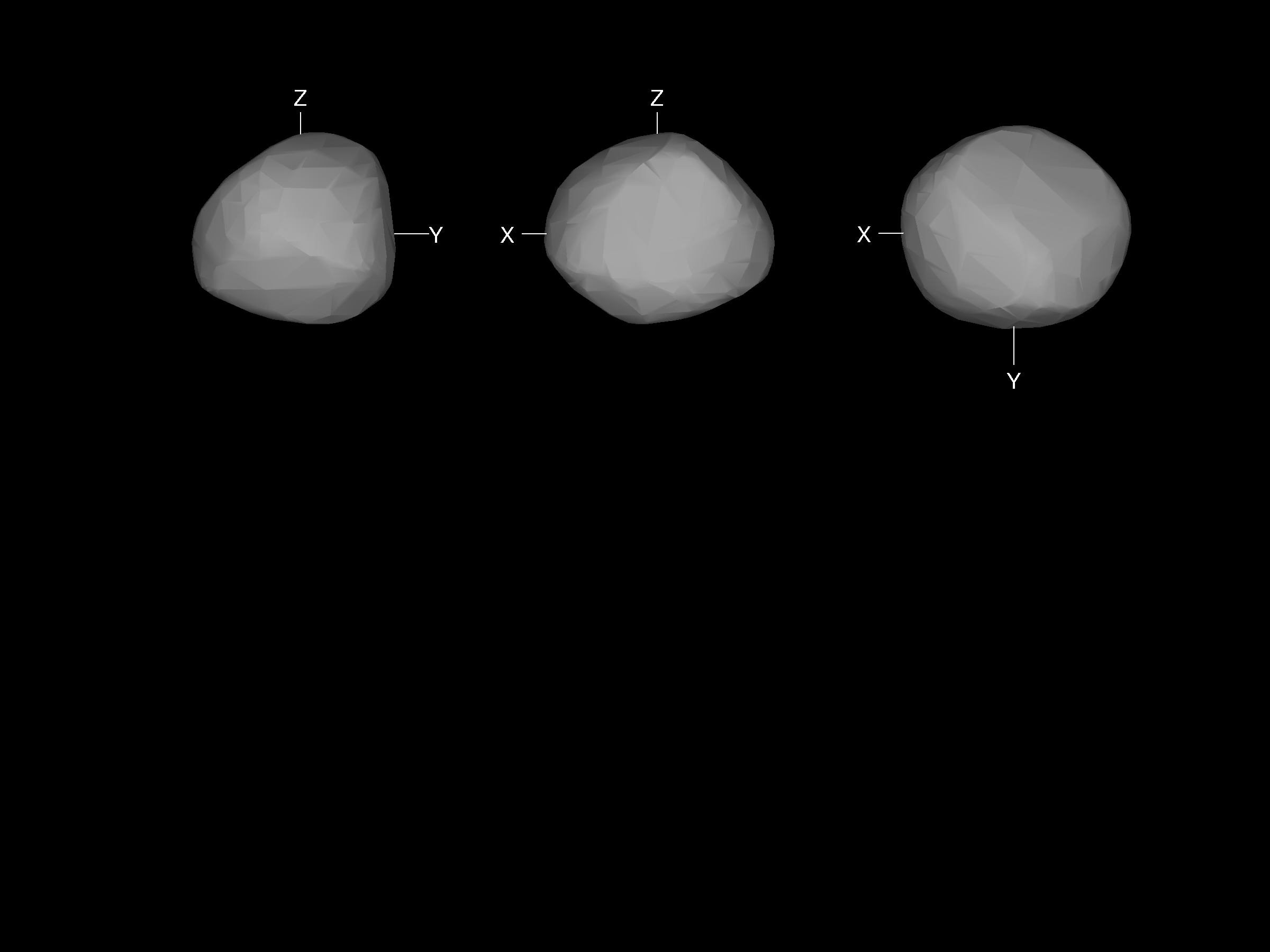}
    \includegraphics[trim=0 7cm 0 0, clip, width=\columnwidth]{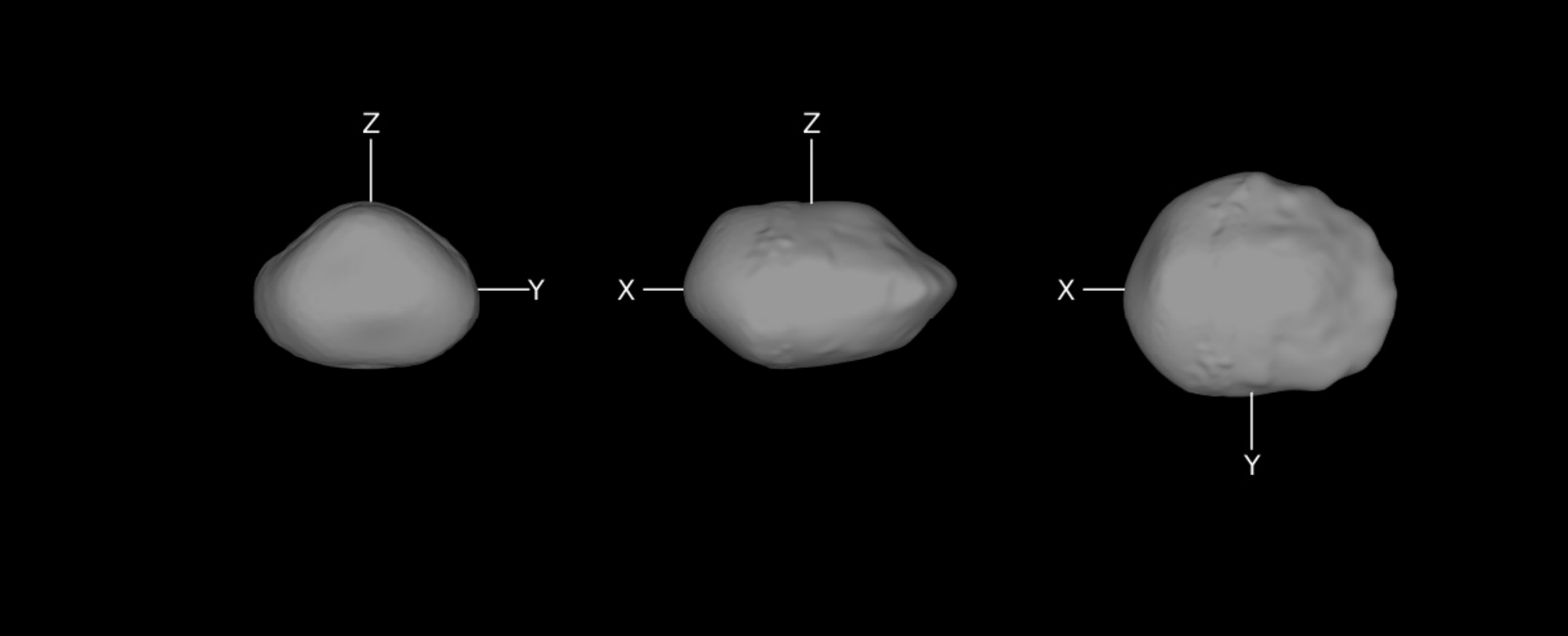}
    \caption{Comparison between the shape of (2867) \v{S}teins reconstructed by our method (top) and that 
reconstructed by \cite{Jor.ea:12} from Rosetta fly-by images and lightcurves (bottom).}
    \label{fig:Steins_shape}
   \end{center}
  \end{figure}

\begin{figure}
   \begin{center}
    \includegraphics[width=\columnwidth]{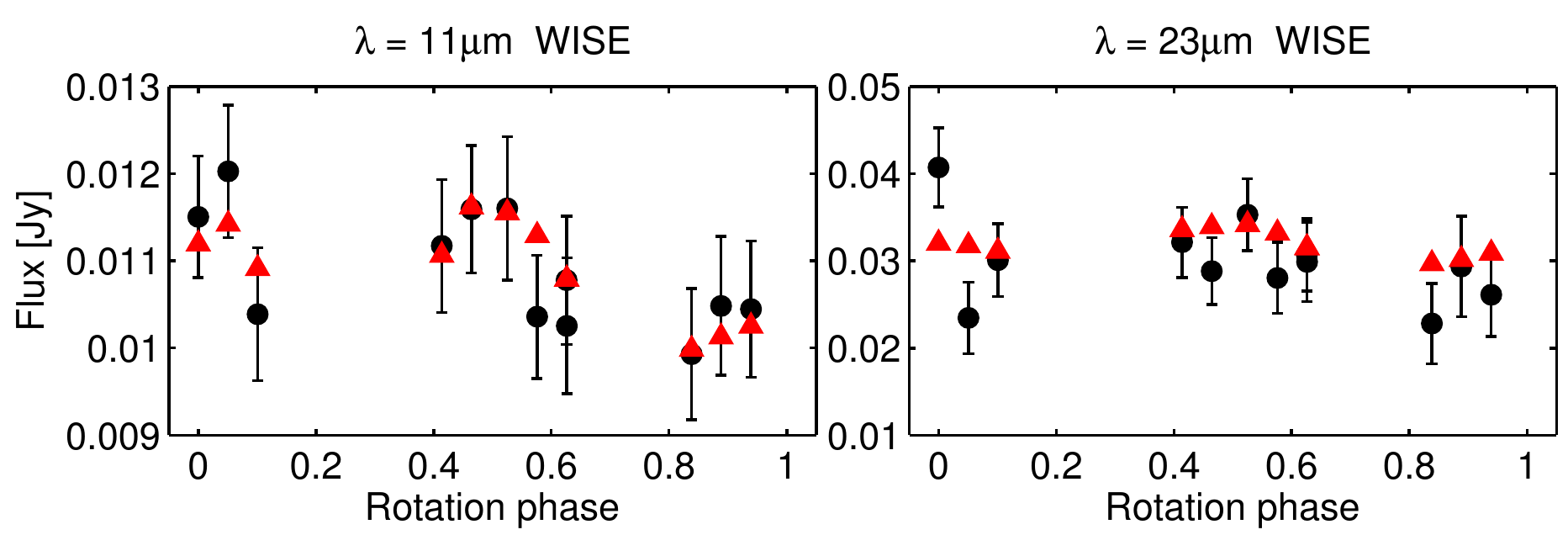}
    \caption{Measured IR fluxes (black circles) of (2867) \v{S}teins and the model fluxes for our model (red triangles).}
    \label{fig:Steins_flux}
   \end{center}
  \end{figure}

Similarly to Lutetia, also for this asteroid we have a detailed shape
model reconstructed from the Rosetta fly-by images and lightcurves
\citep{Jor.ea:12}. But contrary to Lutetia, thermal data for
\v{S}teins are limited to only twelve pairs of WISE data with relative
accuracy of 7\% and 15\% in W3 and W4 filter, respectively. Therefore
the thermal properties are poorly constrained. Models with the
diameter in the range 5.6--6.2\,km and the thermal inertia
70--370\gammaSI\ provide an acceptable fit to both thermal and optical
data. There is a correlation between these two parameters -- solutions
with smaller diameter have also lower thermal inertia. The geometric
visible albedo is in the range 0.4--0.5. One of the possible shape
models is shown in Fig.~\ref{fig:Steins_shape} and the fit to the
thermal data in Fig.~\ref{fig:Steins_flux}. The selected model has an
equivalent diameter of 5.8\,km
and thermal inertia of 200\gammaSI. For
comparison, the shape reconstructed from Rosetta images and
lightcurves has the equivalent diameter $5.26 \pm 0.26$\,km
\citep{Jor.ea:12}. The main difference between the CITPM model and
that of \citep{Jor.ea:12} is that the Rosetta-based model is much more
flat than CITPM model, which might be also the reason why our model
has larger equivalent diameter. \cite{Ley.ea:11} derived from
VIRTIS/Rosetta measurements the thermal inertia $\Gamma = 110 \pm
13$\gammaSI\ for a smooth surface and $\Gamma = 210 \pm
30$\gammaSI\ when a small-scale roughness was
included. \cite{Spj.ea:12} determined the geometric albedo $0.39 \pm
0.02$ at 670\,nm from disk-resolved photometry. This shows that with
low-quality thermal data, the combined inversion is still possible,
but the derived physical parameters have large uncertainties. 

\subsection{(220) Stephania}

\begin{figure}
   \begin{center}
    \includegraphics[width=\columnwidth]{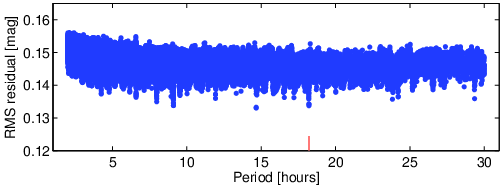}
    \includegraphics[width=\columnwidth]{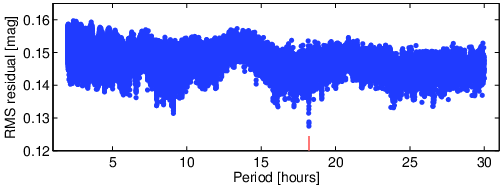}
    \caption{Period search for (220) Stephania for Lowell photometry (top) and Lowell + WISE W1 and W2 data (bottom). 
The correct rotation period of 18.2\,h is marked with a red tick.}
    \label{fig:220_period}
   \end{center}
  \end{figure}

\begin{figure}
   \begin{center}
    \includegraphics[width=\columnwidth]{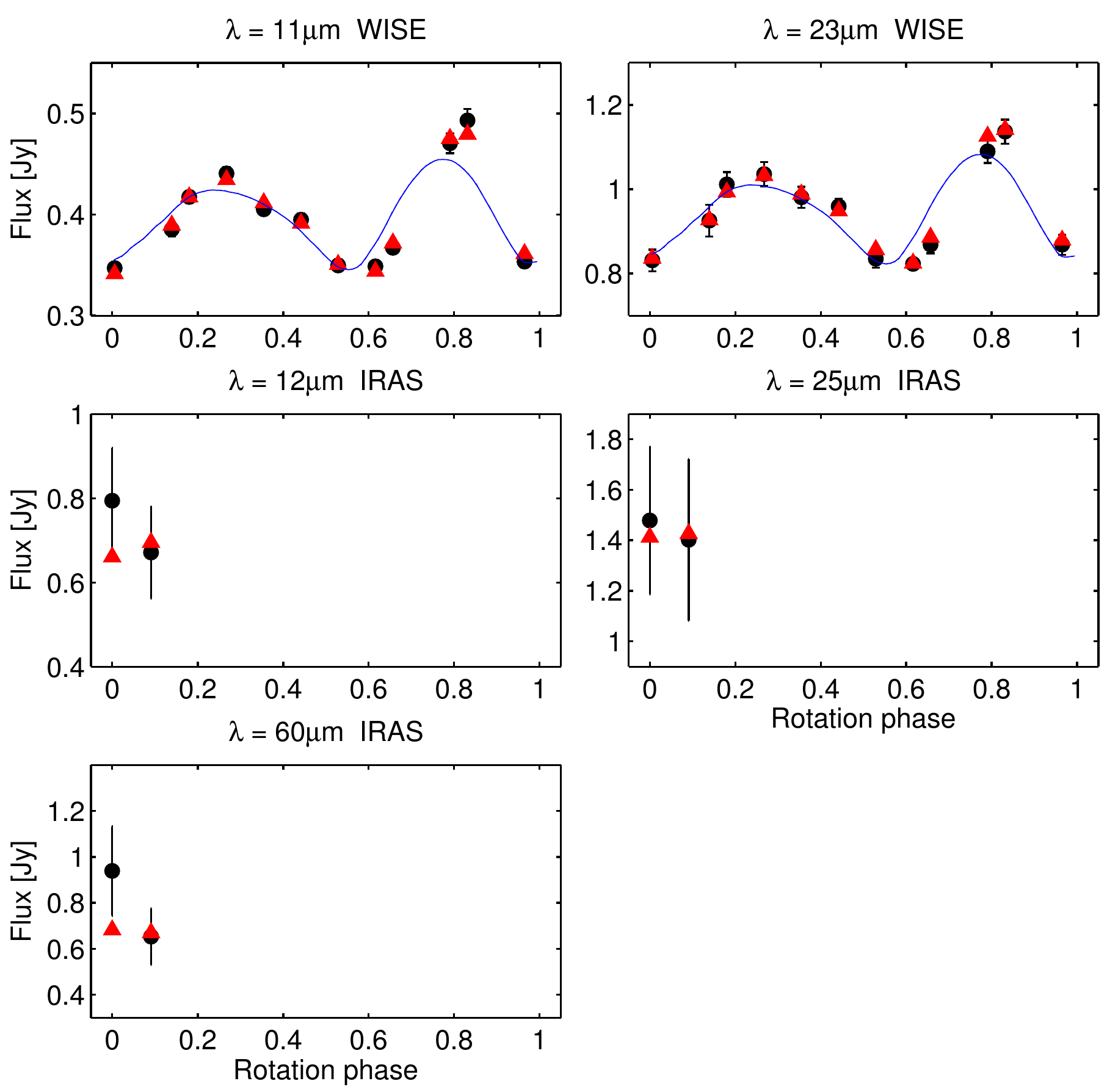}
    \caption{Measured IR fluxes (black circles) of (220) Stephania and the model fluxes for our model (red triangles). 
The WISE data are shown together with the synthetic optical lightcurve (blue) produced by the model.}
    \label{fig:220_flux_fit}
   \end{center}
  \end{figure}

As the last example we selected an asteroid for which optical data are not sufficient to derive a unique model. 
To demonstrate the potential of combining sparse photometry with WISE data, we selected asteroid (220)~Stephania 
that was modelled by \cite{Han.ea:13b} from a set of nine lightcurves and sparse data from the US Naval Observatory 
and the Catalina Sky Survey. The pole direction was $(26^\circ, -50^\circ)$ or $(223, -62^\circ)$. Here we use only 
Lowell sparse photometry and WISE data in W1 (8 points) and W2 (7 points) for the period determination by the lightcurve 
inversion method of \cite{Kaa.ea:01}. As is shown in Fig.~\ref{fig:220_period}, sparse data alone are not sufficient to 
determine the rotation period uniquely. In the periodogram, there are many possible periods (and corresponding models) 
that provide the same level of fit for the data. However, adding WISE data from W1 and W2 filters and assuming that they 
can be treated as reflected light \citep{Dur.ea:16c}, the correct period of about 18.2\,h gives now the global minimum in 
the periodogram. Even if the number of WISE data points is small compared to the Lowell data ($\sim 400$ points), they were 
observed within an interval of one day and their contribution to constraining the period is significant. Then, this period is 
used as a start point for the CITPM model combining now only Lowell photometry, WISE thermal data in W3 and W4 filters, and IRAS 
data. The best model has pole directions $(24^\circ, -60^\circ)$ or $(224^\circ, -59^\circ)$ and thermophysical parameters 
$\Gamma = 15^{+60}_{-10}\,$\gammaSI, $D = 32.2^{+2.0}_{-0.2}$\,km, $p_\text{V} = 0.075 \pm 0.15$. The fit to the IRAS and 
WISE thermal data is shown in Fig.~\ref{fig:220_flux_fit}. The rotation phase shift between the optical and thermal 
lightcurves is very small, only about $10^\circ$, which justifies the use of thermal data for period determination.

\begin{table*}[t]
 \caption{Number of dense lightcurves $N_\text{lc}$, sparse data points $N_\text{sp}$, and IRAS and WISE observations. 
The goodness of fit to thermal data is expressed by the reduced $\chi^2_\text{IR}$}
 \label{tab:data} 
 \centering
 \begin{tabular}{r@{\hspace{1mm}}lccccccccc}
  \hline\hline
  \multicolumn{2}{c}{Asteroid}	& $N_\text{lc}$	& $N_\text{sp}$	& \multicolumn{4}{c}{IRAS}					& \multicolumn{2}{c}{WISE}	& reduced $\chi^2_\text{IR}$	\\
  %\cline{5-8}
  %\cline{9-10}
			& 	&		&	 	& $12\,\mu$m	& $25\,\mu$m	& $60\,\mu$m	& $100\,\mu$m	& $11\,\mu$m	& $22\,\mu$m	&				\\
  \hline
  21 & Lutetia\tablefootmark{a}	& 59		&		& 5		& 5		& 5		& 9		&		&		& 0.80				\\	
  220 & Stephania		&		& 397		& 2		& 2		& 2		&		& 11		& 11		& 0.58				\\
  306 & Unitas			& 15		& 386		& 11		& 11		& 11		& 		& 10		& 13		& 0.68				\\
  2867 & \v{S}teins		& 21		& 614		& 		& 		& 		& 		& 12		& 12		& 0.77				\\

  \hline
 \end{tabular}
 \tablefoot{\tablefoottext{a}{For Lutetia, we did not use sparse data because from the set of 59 dense lightcurves, 
20 were calibrated. The IR data set also contained data from Spitzer and Herschel telescopes.}}
\end{table*}

\begin{table*}[t]
 \caption{Comparison between our CITPM models and independent results. The physical parameters are the ecliptic longitude 
$\lambda$ and latitude $\beta$ of the spin axis direction, the volume-equivalent diameter $D$, the geometric albedo $p_\text{V}$, and the thermal inertia of the surface $\Gamma$.}
 \label{tab:models} 
 \centering
 \begin{tabular}{r@{\hspace{1mm}}lccccccc}
  \hline\hline
  \multicolumn{2}{c}{Asteroid}		& $\lambda$ 		& $\beta$ 		& $D$ 		& $p_\text{V}$			& $\Gamma$	& References		\\
        &				& [deg] 		& [deg]			& [km]		&			& \gammaSI	&			\\
    \hline
  21 	& Lutetia			& 56			& $-7$			& $101 \pm 4$	& 0.19--0.23		& 30--60	& CITPM	($3\sigma$ errors) \\		
	&				& $52.2 \pm 0.4$	& $-7.8 \pm 0.4$	& $98 \pm 2$	& $0.19 \pm 0.01$	&		& \cite{Sie.ea:11}	\\
	&				&			&			& 95.97		&			& $< 10$	& \cite{ORo.ea:12}	\\
	&				&			&			&		&			& $< 30$	& \cite{Kei.ea:12}	\\
	&				&			&			& $98.3\pm 5.9$	& $0.208 \pm 0.025$	& $< 100$	& \cite{Mul.ea:06}	\\
 \hline
  220 	& Stephania			& 24 (or 224)		& $-60$ (or $-59$)	& 32--34	& $0.075 \pm 0.015$	& 5--75		& CITPM			\\
	&				& 26 (or 223)		& $-50$ (or $-62$)	&		&			&		& \cite{Han.ea:13b}	\\
	&				&			&			&$31.738\pm0.219$& $0.069 \pm 0.016$	&		& \cite{Mas.ea:14}	\\
 \hline
  306 	& Unitas			& 82			& $-30$			& $48.7\pm 0.7$	& 0.21--0.27		& 10--100	& CITPM			\\
	&				& 79			& $-35$			& $49 \pm 5$	& 			& 		& \cite{Dur.ea:07, Dur.ea:11}	\\
	&				& 			& 			& 55--56	& 0.14--0.15		& 100--260	& \cite{Del.Tan:09}	\\
	&				&			&			& $47.2\pm0.13$	& $0.201 \pm 0.013$	&		& \cite{Mas.ea:14}	\\
 \hline
  2867	& \v{S}teins			& 142			& $-83$			& 5.6--6.2	& 0.4--0.5		& 70--370	& CITPM			\\
	&				& 94			& $-85 \pm 5$		& $5.26\pm0.26$	& 			& 		& \cite{Jor.ea:12}	\\
	&				&			&			& 		& $0.39 \pm 0.02$	& 		& \cite{Spj.ea:12}	\\ 
	&				&			&			& 		& 			& $210 \pm 30$	& \cite{Ley.ea:11}	\\  
  \hline
 \end{tabular}
\end{table*}

\section{Conclusions and future work}

The new approach of combined inversion of thermal and optical data opens a new possibility to analyze IR data 
and visual photometry for tens of thousands of asteroids for which both data modes are available. As the first 
step, we will apply this method to asteroids for which a shape model exists in the DAMIT database \citep{Dur.ea:10} 
and for which there are WISE or IRAS data,  with the aim to derive complete physical models.

As the next step, WISE data can be processed together with the sparse photometric data with the aim to derive 
unique models in cases when neither visual photometry nor thermal data are sufficient alone. Because WISE data 
constrain the rotation period and Lowell photometry covers various geometries, we expect that thousand of new 
models can be derived from these data sets. 

\begin{acknowledgements}
  The work of J\v{D} was supported by the grant 15-04816S of the Czech
  Science Foundation. 
  BC acknowledges the support of the ESAC Science Faculty for J\v{D} visit.
  VAL has received funding from the European Union's Horizon 2020 Research and Innovation Programme, under Grant Agreement no. 687378.
  VAL and MD acknowledge support from the NEOShield-2 project, which has received funding from the European Union's Horizon 2020 
research and innovation programme under grant agreement no. 640351.
  This publication also makes use of data products
  from NEOWISE, which is a project of the Jet Propulsion
  Laboratory/California Institute of Technology, funded by the
  Planetary Science Division of the National Aeronautics and Space
  Administration.

\end{acknowledgements}

\bibliographystyle{aa} 
\bibliography{bibliography_all} 

\end{document}